\documentclass[12pt,english,aps]{revtex4}
\usepackage[T1]{fontenc}
\usepackage[latin9]{inputenc}
\usepackage[a4paper]{geometry}
\usepackage{amsmath}
\usepackage{graphicx}
\usepackage{amssymb}
\usepackage{esint}

\makeatletter
\@ifundefined{textcolor}{}
{%
 \definecolor{BLACK}{gray}{0}
 \definecolor{WHITE}{gray}{1}
 \definecolor{RED}{rgb}{1,0,0}
 \definecolor{GREEN}{rgb}{0,1,0}
 \definecolor{BLUE}{rgb}{0,0,1}
 \definecolor{CYAN}{cmyk}{1,0,0,0}
 \definecolor{MAGENTA}{cmyk}{0,1,0,0}
 \definecolor{YELLOW}{cmyk}{0,0,1,0}
 }

\makeatother

\usepackage{babel}

\begin{document}

\title{Statistical physics of Bose-Einstein condensed light in a dye microcavity}

\author{Jan Klaers, Julian Schmitt, Tobias Damm, Frank Vewinger, and Martin
Weitz}

\address{Institut f\"ur Angewandte Physik, Universit\"at Bonn, Wegelerstr.
8, 53115 Bonn, Germany}
\begin{abstract}
We theoretically analyze the temperature behavior of paraxial light
in thermal equilibrium with a dye-filled optical microcavity. At low
temperatures the photon gas undergoes Bose-Einstein condensation (BEC),
and the photon number in the cavity ground state becomes macroscopic
with respect to the total photon number. Owing to a grandcanonical
excitation exchange between the photon gas and the dye molecule reservoir,
a regime with unusually large fluctuations of the condensate number
is predicted for this system that is not observed in present atomic
physics BEC experiments.
\end{abstract}
\maketitle
For many problems in statistical physics, one has the freedom of choice
to use different statistical ensembles for their description, as they
often predict the same physical behavior in the thermodynamic limit.
An interesting counterexample (among others) is Bose-Einstein condensation
\cite{Bose,Einstein} where the grandcanonical description leads to
unusually large condensate number fluctuations of order of the total
particle number \cite{Fujiwara,Ziff,Holthaus,Kocharovsky} - in contrast
to the predictions in the (micro-)canonical case that is typically
realized in experiments with ultracold atomic Bose gases \cite{Bongs}.
This peculiar behavior is known as the grandcanonical fluctuation
catastrophe \cite{Holthaus,Kocharovsky}. In recent work, we have
observed Bose-Einstein condensation of a two-dimensional photon gas
in an optical microcavity \cite{Klaers,Klaers2}. Here, the transversal
motional degrees of freedom of the photons are thermally coupled to
the cavity environment by multiple absorption-fluorescence cycles
in a dye medium, with the latter serving both as a heat bath and a
particle reservoir. Due to particle exchange between the photon gas
and the molecular reservoir, grandcanonical experimental conditions
are expected to be realized in this system. 

In this Letter, we discuss the thermalization mechanism and derive
statistical properties of the photon condensate, including its photon
number distribution, fluctuations and intensity correlations. The
main result is that photonic Bose-Einstein condensates, owing to the
grandcanonical nature of the light-matter thermalization, can show
unusually large particle number fluctuations, which are not observed
in present atomic Bose-Einstein condensates. Our calculations are
done in the limit of a non-interacting photon gas.

The system under investigation, as shown in Fig. 1a, consists of a
microresonator formed by two perfectly reflecting spherically curved
mirrors enclosing a dye medium. In the cavity, optical photons are
permanently absorbed and re-emitted by dye molecules. For a sufficiently
small mirror spacing, this process will maintain the longitudinal
mode number of the photons denoted by $q$, see \cite{Klaers,Klaers2,Klaers3}
for a detailed description of the experiment. The photon gas effectively
becomes two-dimensional, as only the two transversal degrees of freedom
remain. Throughout this Letter, we do not consider experimental imperfections
as mirror losses or non-radiative decay of dye excitations. Moreover,
we do not consider coherent time evolution of the combined light-dye
system (strong coupling) which would result in polaritonic eigenstates.
Under the experimental conditions of \cite{Klaers,Klaers2}, this
is inhibited by a rapid, collision-induced decoherence process.

\begin{figure}[t]
\begin{centering}
\includegraphics[width=8cm]{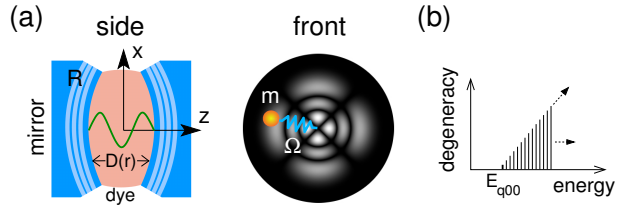}
\par\end{centering}

\vspace{-1mm}

\caption{(a) Optical microresonator enclosing a dye medium (left). For a fixed
longitudinal mode number $q$, the photon gas can be mapped onto a
two-dimensional gas of massive particles (effective mass $m$) confined
in a harmonic trap with trapping frequency $\Omega$ (right) (b) Scheme
of the density of states in the resonator.}

\vspace{-4mm}

\end{figure}
There is a close connection to the statistical physics of the atomic
two-dimensional Bose gas, which can be seen by expressing the photon
energy as a function of the longitudinal ($k_{z}$) and transversal
wave number ($k_{r}$), $E\hspace{-1mm}=\hspace{-1mm}\hbar\tilde{c}\sqrt{k_{z}^{2}\hspace{-1mm}+\hspace{-1mm}k_{r}^{2}}$,
with $\tilde{c}$ as the speed of light in the medium. The boundary
condition in the $z$-direction is incorporated by an ansatz $k_{z}(r)\hspace{-1mm}=\hspace{-1mm}q\pi/D(r)$,
where $D(r)\hspace{-1mm}=\hspace{-1mm}D_{0}\hspace{-1mm}-\hspace{-1mm}2(R\hspace{-1mm}-\hspace{-1mm}\sqrt{R^{2}\hspace{-1mm}-\hspace{-1mm}r^{2}})$
is the mirror separation at a distance $r$ from the optical axis
and $R$ is the radius of curvature. In a paraxial approximation,
with $k_{r}\ll k_{z}(0)$ and $r\ll R$, this yields \cite{Klaers3}\begin{eqnarray}
E & \simeq & m\tilde{c}^{2}+\frac{(\hbar k_{r})^{2}}{2m}+\frac{m\Omega^{2}}{2}\, r^{2}\:\textrm{,}\end{eqnarray}
which is formally equivalent to the energy of a two-dimensional harmonic
oscillator when defining a photon mass $m\hspace{-1mm}=\hspace{-1mm}\hbar k_{z}(0)/\tilde{c}$
and a trapping frequency $\Omega\hspace{-1mm}=\hspace{-1mm}\tilde{c}/\sqrt{D_{0}R/2}$.
The photon gas can thus be mapped onto a (2d) gas of massive bosons
in a harmonic trap \nobreakdash- a system known to undergo BEC at
a non-zero temperature \cite{Bagnato,Haugset}. 

\textit{Thermal equilibrium and condensation.} \nobreakdash- First
we discuss the thermalization process in more detail. Thermal equilibrium
between photons and dye molecules will be shown to arise under two
conditions: (i) the Einstein coefficients of the dye medium fulfill
the Kennard-Stepanov law and (ii) chemical equilibrium between photons,
excited and ground state molecules is achieved. 

(i) The Kennard-Stepanov law \cite{Ross,McCumber} (and references
therein) relates the Einstein coefficients of stimulated absorption
and emission, and can be stated in the form\begin{equation}
\frac{B_{21}(\omega)}{B_{12}(\omega)}=\frac{w_{\downarrow}}{w_{\uparrow}}\, e^{-\frac{\hbar(\omega-\omega_{0})}{k_{\textrm{B}}T}}\:\textrm{,}\label{eq:KS}\end{equation}
where $B_{12,21}(\omega)$ are the Einstein coefficients of absorption
and emission, $\omega_{0}$ is the zero-phonon line of the dye and
$w_{\downarrow,\uparrow}=\int_{\epsilon\ge0}\mathcal{D}_{\downarrow,\uparrow}(\epsilon)\exp\left(-\epsilon/k_{\text{B}}T\right)\, d\epsilon$
are statistical weights related to the rovibronic density of states
$\mathcal{D}_{\downarrow,\uparrow}(\epsilon)$ of ground ($\downarrow$)
and excited ($\uparrow$) dye state. This law is often fulfilled for
dyes in liquid solution and goes back to a thermalization process
of the rovibronic dye state due to frequent collisions with solvent
molecules \cite{Supplemental}. 

(ii) The excitation exchange between photon gas and molecules can
be seen as a photochemical reaction of the type $\gamma+\hspace{-1mm}\downarrow\:\leftrightharpoons\:\uparrow$.
Here $\gamma$ stands for a photon, $\downarrow$ for a ground state
molecule, and $\uparrow$ for an excited molecule. The energy of optical
photons is well above thermal energy at room temperature, i.e. $E\hspace{-1mm}\gg\hspace{-1mm}k_{\text{B}}T$,
and the conversion of ground state molecules into excited molecules
by thermal fluctuations is thus negligible. In this situation, the
total number of photons and excited molecules is not adjusted by temperature
as for black-body radiation, but is a conserved quantity (it depends
on the initial state of the photon gas and the molecular medium).
The chemical potential of the photons $\mu_{\gamma}$ then can become
non-zero. In equilibrium, $\mu_{\gamma}$ is related to the other
chemical potentials by $\mu_{\gamma}+\mu_{\downarrow}=\mu_{\uparrow}$
\cite{Wuerfel}. Starting from this, the photon chemical potential
$\mu_{\gamma}$ can be shown \cite{Supplemental} to be related to
the excitation level $\rho_{\uparrow}/\rho_{\downarrow}$ in the dye
medium by\begin{equation}
e^{\frac{\mu_{\gamma}}{k_{\textrm{B}}T}}=\frac{w_{\downarrow}}{w_{\uparrow}}\,\frac{\rho_{\uparrow}}{\rho_{\downarrow}}\, e^{\frac{\hbar\omega_{0}}{k_{\text{B}}T}}\enskip\textrm{.}\label{eq:z2}\end{equation}
Here $\rho_{\downarrow,\uparrow}$ are the densities of ground and
excited state molecules. The chemical potential $\mu_{\gamma}$ is
spatially homogeneous in equilibrium (as is the temperature), from
which follows that the excitation level of the medium $\rho_{\uparrow}/\rho_{\downarrow}$
is also position-independent.

The thermalization process is considered as a random walk in the configuration
space of all allowed light field states. Here, a state $K$ is given
by the cavity mode occupation numbers $K=(n_{0}^{K},n_{1}^{K},n_{2}^{K},\ldots)$.
The mode occupation numbers are frequently altered by photon absorption
and emission processes. In first order perturbation theory, the rates
(per volume) for absorption and emission of one photon in mode $i$
at cavity position $\mathbf{r}$, denoted by $R_{12}^{K,i}(\mathbf{r})$
and $R_{21}^{K,i}(\mathbf{r})$, have the form \begin{eqnarray}
R_{12}^{K,i}(\mathbf{r}) & = & B_{12}(\omega_{i})\, u^{i}(\mathbf{r})\,\rho_{\downarrow}\, n_{i}^{K}\label{eq:R12}\\
R_{21}^{K,i}(\mathbf{r}) & = & B_{21}(\omega_{i})\, u^{i}(\mathbf{r})\,\rho_{\uparrow}\,(n_{i}^{K}+1)\:\textrm{,}\label{eq:R21}\end{eqnarray}
where $u^{i}(\mathbf{r})$ is the spectral energy density of one photon
in mode $i$. We first assume a grandcanonical ensemble limit, i.e.
we consider the number of dye molecules as sufficiently large, that
the change of the excitation level $\rho_{\uparrow}/\rho_{\downarrow}$
from photon absorption and emission can be neglected. Thus, the species
densities $\rho_{\downarrow,\uparrow}$ are treated as fixed parameters. 

In the theory of random walks, a well known detailed balance criterion
exists that determines if the rates given by eq. \eqref{eq:R12} and
\eqref{eq:R21} lead to equilibrium \cite{Landau,Norman}. Suppose
that a state $K'$ emerges from state $K$ by the absorption of a
photon in mode $i$, with $n_{i}^{K'}=n_{i}^{K}-1$, and accordingly
that $K$ emerges from $K'$ by an emission process into this mode.
The corresponding (local) rates are $R_{12}^{K,i}(\mathbf{r})$ and
$R_{21}^{K',i}(\mathbf{r})$; and with eq. \eqref{eq:R12} and \eqref{eq:R21},
their ratio is given by $R_{12}^{K,i}(\mathbf{r})/R_{21}^{K',i}(\mathbf{r})\hspace{-0.7mm}=\hspace{-0.7mm}B_{12}(\omega_{i})\,\rho_{\downarrow}/B_{21}(\omega_{i})\,\rho_{\uparrow}$.
Thermal equilibrium will be reached, if this ratio is given by the
Boltzmann factor of the energy difference between $K$ and $K'$ \cite{Norman},
i.e.\begin{equation}
\frac{B_{12}(\omega_{i})}{B_{21}(\omega_{i})}\,\frac{\rho_{\downarrow}}{\rho_{\uparrow}}=e^{\frac{\hbar\omega_{i}-\mu_{\gamma}}{k_{\textrm{B}}T}}\enskip\textrm{.}\label{eq:DB}\end{equation}
If one now applies the Kennard-Stepanov relation, eq. \eqref{eq:KS},
and assumes chemical equilibrium, eq. \eqref{eq:z2}, the detailed
balance condition eq. \eqref{eq:DB} is indeed verified. Thus, the
state of the photon gas will thermalize to the temperature of the
dye solution $T$ at a chemical potential $\mu_{\gamma}$ related
to the excitation level of the dye molecules. In particular, the average
occupation number of mode $i$, denoted by $\bar{n}_{i}$, can be
determined by balancing the average absorption and emission rates
at a given cavity position. This gives the expected Bose-Einstein
distribution $\bar{n}_{i}\hspace{-0.7mm}=\hspace{-0.7mm}\bigl(\exp\left[\left(\hbar\omega_{i}\hspace{-0.7mm}-\hspace{-0.7mm}\mu_{\gamma}\right)\hspace{-0.7mm}/k_{\textrm{B}}T\right]\hspace{-0.7mm}-\hspace{-0.7mm}1\bigr)^{-1}$.
For atomic Bose gases, one typically omits the rest energy $mc^{2}$
from the Hamiltonian. We will adopt this convention for the photon
gas by removing the ground state energy $E_{q00}$ from all energies
including the chemical potentials. With $\mu\hspace{-0.7mm}=\hspace{-0.7mm}\mu_{\gamma}\hspace{-0.7mm}-\hspace{-0.7mm}E_{q00}$,
the Bose-Einstein distribution becomes\begin{equation}
\bar{n}(u)=\frac{g(u)}{e^{\frac{u-\mu}{k_{\textrm{B}}T}}-1}\enskip\textrm{,}\label{eq:BEdist2}\end{equation}
where $u\hspace{-0.7mm}=\hspace{-0.7mm}E\hspace{-0.7mm}-\hspace{-0.7mm}E_{q00}\hspace{-0.7mm}=\hspace{-0.7mm}0,\hbar\Omega,2\hbar\Omega,\ldots$
is the reduced photon energy, and we have used the degeneracy $g(u)\hspace{-0.7mm}=\hspace{-0.7mm}u/\hbar\Omega\hspace{-0.7mm}+\hspace{-0.7mm}1$
to combine all modes of same energy, see Fig. 1b. For the sake of
simplicity, we neglect the polarization degeneracy of each mode throughout
this Letter, which would give an additional degeneracy factor of $2$.
This case is discussed in the Supplemental Material \cite{Supplemental}.
For this system, a phase transition is expected at a critical temperature
of \cite{Bagnato,Haugset} \begin{equation}
T_{\textrm{c}}=\frac{\sqrt{6}\hbar\Omega}{\pi k_{\textrm{B}}}\sqrt{\bar{N}}=\frac{\sqrt{12}\hbar\tilde{c}}{\pi k_{\textrm{B}}}\sqrt{\frac{1}{D_{0}}\frac{\bar{N}}{R}}\enskip\textrm{,}\label{eq:TBEC}\end{equation}
at which the ground state occupation $\bar{n}_{0}\equiv\bar{n}(0)$
becomes a macroscopic fraction of the total average photon number
$\bar{N}$.

\textit{Condensate fluctuations.} \nobreakdash- For grandcanonical
Bose-Einstein condensation, i.e. in the presence of an infinitely
large particle reservoir, condensate number fluctuations of order
of the total particle number occur \cite{Fujiwara,Ziff,Holthaus,Kocharovsky}.
However, no excess fluctuations occur in the (micro-)canonical ensemble,
which is usually realized in atomic BEC experiments \cite{Bongs}.
For the photon Bose-Einstein condensate, the ground state mode is
coupled to the electronic transitions of a given number of dye molecules.
In this way, the condensate exchanges excitations with a reservoir
of size $M$, which is given by the product of dye concentration and
ground state mode volume. 

We start with the master equation for the probability $p_{n}=p_{n}(t)$
to find $n(\equiv n_{0})$ photons in the ground state at time $t$.
The flow of probability between the ground mode and its reservoir
is governed by \cite{Abraham} \begin{equation}
\dot{p}_{n}=R_{n\hspace{-0.4mm}-\hspace{-0.4mm}1}^{21}\hspace{0.4mm}p_{n\hspace{-0.4mm}-\hspace{-0.4mm}1}-(R_{n}^{12}\hspace{-0.7mm}+\hspace{-0.7mm}R_{n}^{21})\hspace{0.4mm}p_{n}+R_{n\hspace{-0.4mm}+\hspace{-0.4mm}1}^{12}\hspace{0.4mm}p_{n\hspace{-0.4mm}+\hspace{-0.4mm}1}\enskip\textrm{,}\label{eq:master}\end{equation}
where $R_{n}^{12}\hspace{-0.7mm}=\hspace{-0.7mm}\hat{B}_{12}(M\hspace{-0.7mm}-\hspace{-0.7mm}X\hspace{-0.7mm}+\hspace{-0.7mm}n)n$
is the rate of absorption and $R_{n}^{21}\hspace{-0.7mm}=\hspace{-0.7mm}\hat{B}_{21}(X\hspace{-0.7mm}-\hspace{-0.7mm}n)(n\hspace{-0.7mm}+\hspace{-0.7mm}1)$
the rate of stimulated and spontaneous emission for a configuration
with $n$ photons, $X\hspace{-0.7mm}-\hspace{-0.7mm}n$ electronically
excited molecules and $M\hspace{-0.7mm}-\hspace{-0.7mm}X\hspace{-0.7mm}+\hspace{-0.7mm}n$
ground state molecules. The excitation number $X$ is the sum of the
ground mode photon number and the number of excited molecules in the
reservoir. In this calculation, $X$ is treated as a constant \cite{Supplemental},
i.e. it is not expected to perform large fluctuations on its own.
The rates $R_{n}^{12}$ and $R_{n}^{21}$ follow from eq. \eqref{eq:R12},
\eqref{eq:R21} by integrating over volume, and setting $\hat{B}_{12,21}\hspace{-0.7mm}:=\hspace{-0.7mm}B_{12,21}\hspace{-0.35mm}(\hspace{-0.25mm}E_{q00}\hspace{-0.25mm}/\hspace{-0.25mm}\hbar)\, u^{q00}(\mathbf{0})$,
where $u^{q00}(\mathbf{0})$ is the spectral energy density of the
ground mode on the optical axis. For large times, $p_{n}(t)$ is expected
to become stationary, $\dot{p}_{n}(\infty)\hspace{-0.7mm}=\hspace{-0.7mm}0$,
and approach its equilibrium value $\mathcal{P}_{n}\hspace{-0.7mm}:=\hspace{-0.7mm}p_{n}(\infty)$.
In this asymptotic case, the solution of the master equation eq. \eqref{eq:master}
is found to be $\mathcal{P}_{n}=\mathcal{P}_{0}\prod_{k=0}^{n-1}R_{k}^{21}/R_{k+1}^{12}$,
which is\begin{eqnarray}
\frac{\mathcal{P}_{n}}{\mathcal{P}_{0}} & = & \frac{\left(M\hspace{-0.7mm}-\hspace{-0.7mm}X\right)!\, X!}{\left(M\hspace{-0.7mm}-\hspace{-0.7mm}X\hspace{-0.7mm}+\hspace{-0.7mm}n\right)!\,\left(X\hspace{-0.7mm}-\hspace{-0.7mm}n\right)!}\,\left(\frac{\hat{B}_{21}}{\hat{B}_{12}}\right)^{n}\textrm{.}\label{eq:Pn}\end{eqnarray}
\begin{figure}[t]
\vspace{-3mm}

\begin{centering}
\hspace{0mm}\includegraphics[width=8.35cm]{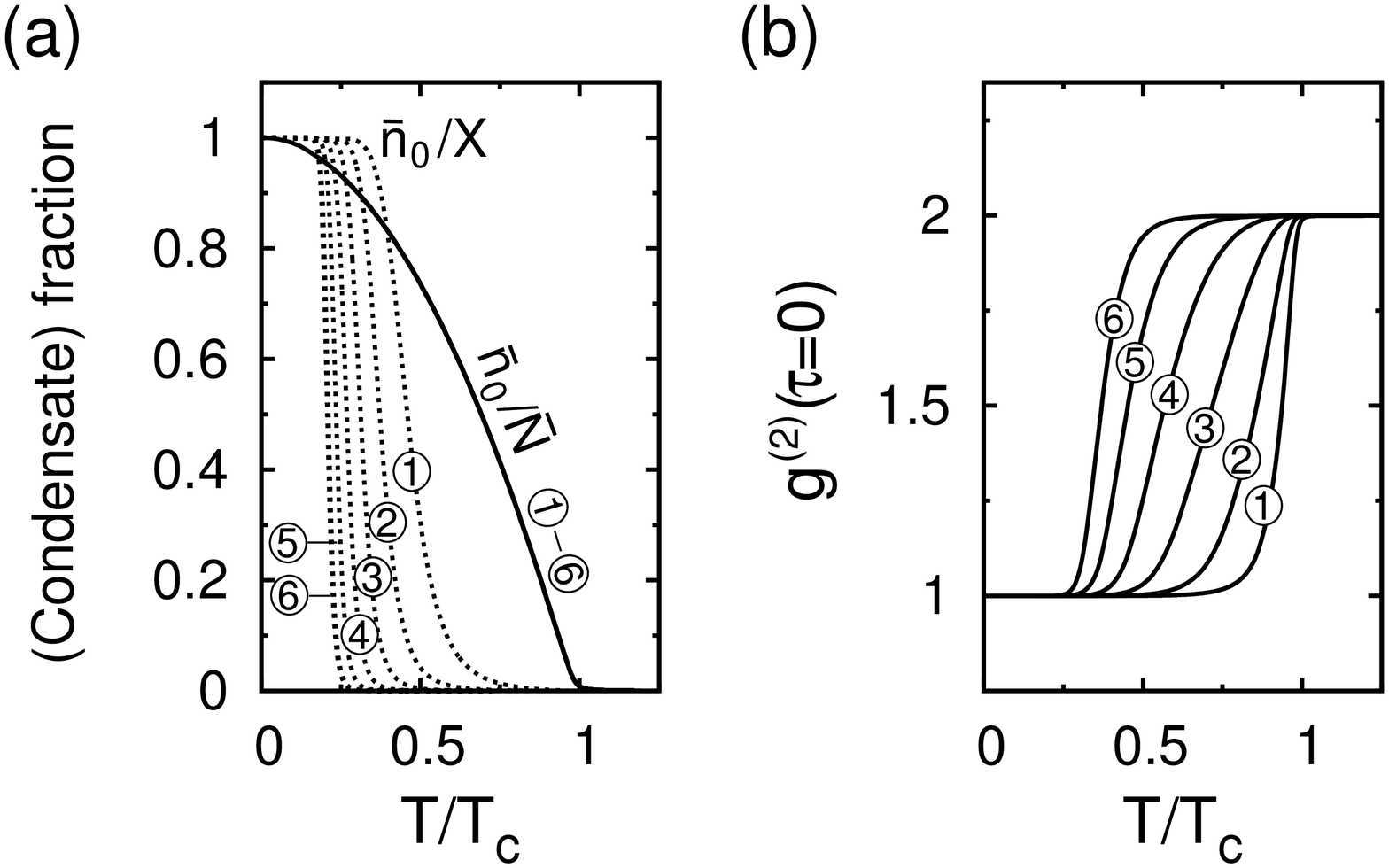}\hspace{0mm}
\par\end{centering}

\vspace{-10mm}

\caption{(a) Condensate fraction $\bar{n}_{0}/\bar{N}$ (solid lines) and fraction
$\bar{n}_{0}/X$ (dashed lines) as a function of the reduced temperature
$T/T_{\textrm{c}}$ for a photon number of $\bar{N}=10^{4}$, six
reservoir sizes from $M_{1}\hspace{-0.7mm}=\hspace{-0.7mm}10^{8}$,
$M_{2}\hspace{-0.7mm}=\hspace{-0.7mm}10^{9}$ up to $M_{6}=10^{13}$
and a dye-cavity detuning of $\hbar\Delta/k_{\text{B}}T_{\text{c}}\hspace{-0.7mm}=\hspace{-0.7mm}-4.35$.
(b) Normalized condensate fluctuations $g^{(2)}(0)\hspace{-0.7mm}=\hspace{-0.7mm}\left\langle n_{0}\,(n_{0}\hspace{-0.7mm}-\hspace{-0.7mm}1)\right\rangle /\bar{n}_{0}^{2}$
as a function of $T/T_{\textrm{c}}$ for various $M$ (as in (a)),
revealing large number fluctuations even far below $T_{\text{c}}$.}

\vspace{2mm}

\begin{centering}
\hspace{-3mm}\includegraphics[width=8cm]{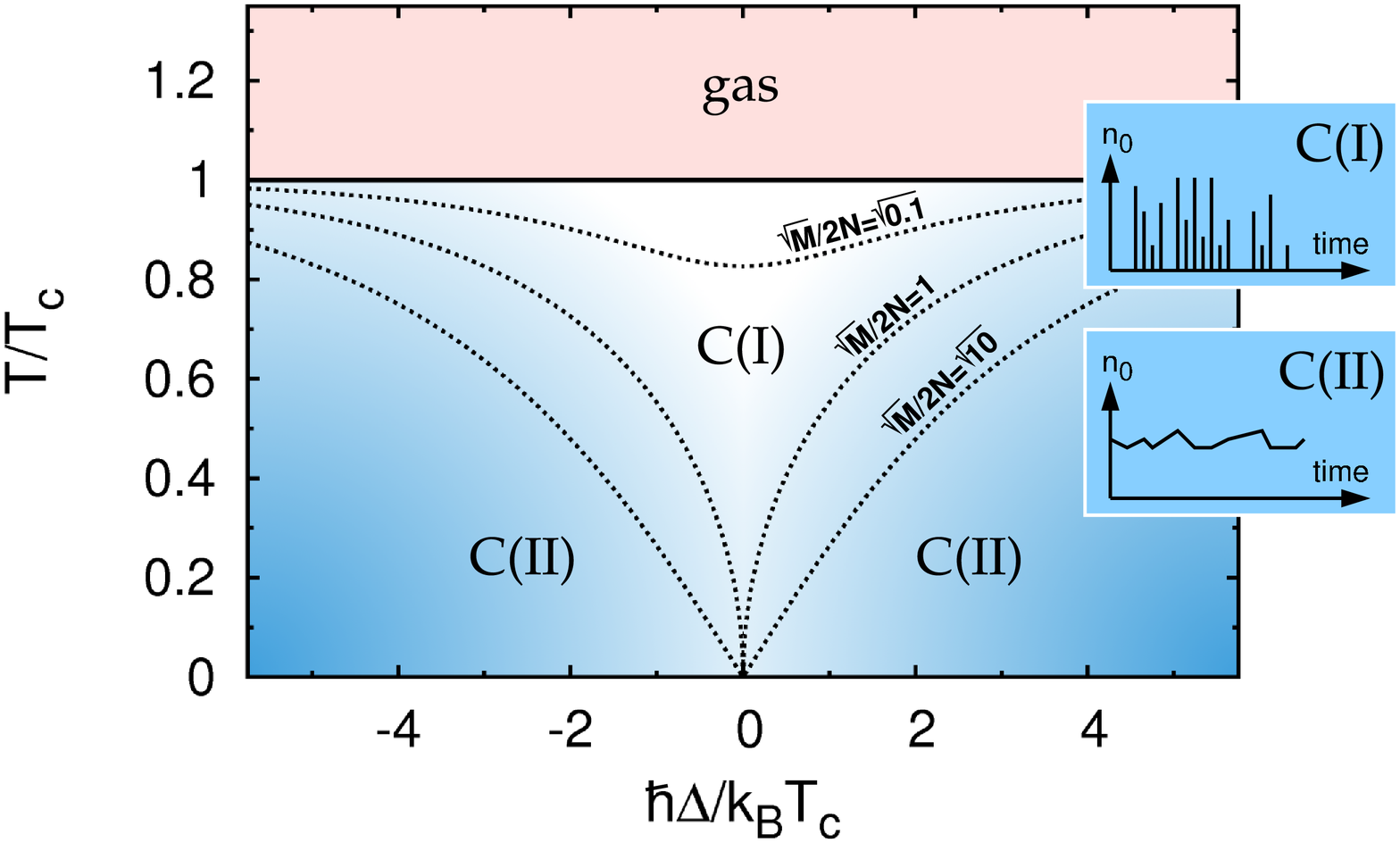}
\par\end{centering}

\vspace{-4mm}

\caption{Phase diagram of the two-dimensional photon gas for fixed average
photon number $\bar{N}$, as a function of the reduced temperature
$T/T_{\textrm{c}}$ and the dye-cavity detuning $\hbar\Delta/k_{\textrm{B}}T_{\textrm{c}}$.
The solid line marks the BEC phase transition. The dashed lines (three
cases are shown) separate two regimes: a condensate with large number
fluctuations and a BE-like photon number distribution C(I), and a
non-fluctuating condensate obeying Poisson statistics C(II). The temperature
of the crossover C(I)-C(II) depends on the ratio $\sqrt{M}/\bar{N}$,
with the reservoir size $M$ as the number of dye molecules in the
mode volume of the ground state. The insets give a sketch of the corresponding
temporal evolution of the condensate photon number $n_{0}(t)$.}

\vspace{-3mm}

\end{figure}
This photon number distribution can be used to obtain both the average
ground state occupation and its fluctuations. We consider experimental
conditions, where the temperature of the system is varied, with the
total average photon number $\bar{N}$ (in all modes) being fixed.
Correspondingly, upon temperature variations the excitation level
$\rho_{\uparrow}/\rho_{\downarrow}$ of the dye medium and the total
ground mode excitation parameter $X$ have to be readjusted (see Supplemental
Material \cite{Supplemental}). 

Figure 2a shows the ground mode occupation $\bar{n}_{0}/\bar{N}$
(solid lines) as a function of the reduced temperature $T/T_{\textrm{c}}$
for a constant average photon number of $\bar{N}\hspace{-0.7mm}=\hspace{-0.7mm}10^{4}$,
six reservoir sizes from $M_{1}\hspace{-0.7mm}=\hspace{-0.7mm}10^{8}$
over $M_{2}\hspace{-0.7mm}=\hspace{-0.7mm}10^{9}$ up to $M_{6}\hspace{-0.7mm}=\hspace{-0.7mm}10^{13}$,
and a dye-cavity detuning of $\hbar\Delta/k_{\text{B}}T_{\text{c}}\hspace{-0.7mm}:=\hspace{-0.7mm}(E_{q00}\hspace{-1mm}-\hspace{-1mm}\hbar\omega_{0})/k_{\text{B}}T_{\text{c}}\hspace{-0.7mm}=\hspace{-0.7mm}-4.35$,
which adjusts $\hat{B}_{21}/\hat{B}_{12}$ via eq. \ref{eq:KS} and
is experimentally realizable \cite{Klaers,Klaers2}. Upon reaching
criticality, the ground mode occupation becomes a macroscopic fraction
of the total photon number. The occupation level closely follows the
analytic solution $\bar{n}_{0}/\bar{N}\hspace{-0.7mm}=\hspace{-0.7mm}1\hspace{-0.7mm}-\hspace{-0.7mm}(T/T_{\textrm{c}})^{2}$
(not shown in Fig. 2a), nearly independent of the reservoir size $M$
(the solid lines for different $M$ essentially overlay in Fig. 2a).
The dashed lines shows the fraction $\bar{n}_{0}/X$ of ground mode
occupation and total excitation number versus temperature, which reveals
that the vast majority of excitations is stored as electronic excitations
in the medium down to relatively low temperatures. Figure 2b gives
the normalized ground mode fluctuations in terms of the zero-delay
autocorrelation function $g^{(2)}(0)\hspace{-0.7mm}=\hspace{-0.7mm}\bigl\langle n_{0}\,(n_{0}\hspace{-0.7mm}-\hspace{-0.7mm}1)\bigr\rangle/\bar{n}_{0}^{2}$
versus temperature. For $T\hspace{-0.7mm}\ge\hspace{-0.7mm}T_{c}$,
one finds the usual case of strong intensity fluctuations with $g^{(2)}(0)\hspace{-0.7mm}=\hspace{-0.7mm}2$,
accompanied by a Bose-Einstein-like photon number distribution. Interestingly,
for large reservoir sizes $M$ the intensity correlation function
remains at this value even at temperatures below $T_{\text{c}}$,
i.e. when condensation sets in. In this regime, large condensate number
fluctuations occur due to grandcanonical particle exchange with the
molecular reservoir. At even lower temperatures, $T\ll T_{\text{c}}$,
the fluctuations are damped, and one finds $g^{(2)}(0)\hspace{-0.7mm}\simeq\hspace{-0.7mm}1$,
accompanied by a Poissonian photon statistics. We have no indication
that the crossover from Bose-Einstein to Poissonian statistics is
accompanied by an additional phase transition, i.e. a non-analytic
behavior of thermodynamic functions.

In general, the photon number distribution eq. \eqref{eq:Pn} can
either be Bose-Einstein-like, Poisson-like or an intermediate case.
It will be Bose-Einstein-like, if the molecular reservoir is so large
that the ratio $\mathcal{P}_{n+1}/\mathcal{P}_{n}=(\hat{B}_{21}/\hat{B}_{12})\,(X\hspace{-0.7mm}-\hspace{-0.7mm}n)/(M\hspace{-0.7mm}-\hspace{-0.7mm}X\hspace{-0.7mm}+\hspace{-0.7mm}n\hspace{-0.7mm}+\hspace{-0.7mm}1)$
is approximately constant for all relevant photon numbers $n$. Then
$\mathcal{P}_{n}$ has its peak value at $n\hspace{-0.7mm}=\hspace{-0.7mm}0$
and decays exponentially like a geometric series, with $\mathcal{P}_{n}/\mathcal{P}_{0}\simeq[(\hat{B}_{21}/\hat{B}_{12})\, X/(M\hspace{-0.7mm}-\hspace{-0.7mm}X)]^{n}$.
Thus, the most probable event is to find no photons at all. Note that
this can be achieved by increasing $M$, while keeping the excitation
level $\rho_{\uparrow}/\rho_{\downarrow}\simeq X/(M\hspace{-0.7mm}-\hspace{-0.7mm}X)$
fixed (which maintains $\mu$, $\bar{n}_{0}$, $\bar{N}$). For Poissonian
statistics, $\mathcal{P}_{n}$ has its maximum at a non-zero $n$.
Due to the smooth crossover the distinction between these two statistical
regimes is not unambiguous. However, a natural choice is to consider
the point at which 'finding zero photons' ceases to be the most probable
event. This occurs at $\mathcal{P}_{0}\hspace{-0.5mm}=\hspace{-0.5mm}\mathcal{P}_{1}$
(compare also to the laser threshold described in \cite{Scully})
and with eq. \eqref{eq:Pn} corresponds to $(M\hspace{-0.7mm}+\hspace{-0.7mm}1)/X\hspace{-0.7mm}=\hspace{-0.7mm}1\hspace{-0.7mm}+\hspace{-0.7mm}\hat{B}_{21}/\hat{B}_{12}$.
The temperature $T_{\text{x}}$ at which this occurs, for fixed $\bar{N}$
and $M$, can be shown to fulfill the transcendental equation\begin{equation}
\bar{N}\hspace{-0.7mm}-\hspace{-0.7mm}\frac{\pi}{6}\left(\frac{k_{B}T_{\text{x}}}{\hbar\Omega}\right)^{2}\hspace{-0.7mm}\simeq\sqrt{\frac{M}{\bigl(1\hspace{-0.5mm}+\hspace{-0.5mm}e^{\frac{\hbar\Delta}{k_{\textrm{B}}T_{\text{x}}}}\bigr)\bigl(1\hspace{-0.5mm}+\hspace{-0.5mm}e^{-\frac{\hbar\Delta}{k_{\textrm{B}}T_{\text{x}}}}\bigr)}}\enskip\textrm{.}\label{eq:Tx}\end{equation}
For zero dye-cavity detuning $\Delta\hspace{-0.7mm}=\hspace{-0.7mm}0$,
one finds the analytic solution $T_{\textrm{x},\Delta\hspace{-0.5mm}=\hspace{-0.5mm}0}\simeq T_{\textrm{c}}\,\sqrt{1\hspace{-0.5mm}-\hspace{-0.5mm}\sqrt{M}/2\bar{N}}$,
provided that $\sqrt{M}/2\bar{N}\hspace{-0.7mm}<\hspace{-0.7mm}1$,
while for general detunings $\Delta$, eq. \eqref{eq:Tx} has to be
solved numerically. Figure 3 gives a phase diagram, where solutions
for three different cases $\sqrt{M}/2\bar{N}\hspace{-0.5mm}=\hspace{-0.5mm}\sqrt{0.1},1,\sqrt{10}$
are marked as dashed lines, which separate two regimes of the condensate,
denoted by C(I) with Bose-Einstein-like photon statistics and C(II)
with Poisson statistics, respectively. In terms of second order correlations,
the dashed lines correspond to $g^{(2)}(0)\hspace{-0.7mm}\simeq\hspace{-0.7mm}1.571$.
Note that both $T_{\text{c}}$ and $T_{\text{x}}$ are conserved in
a thermodynamic limit $\bar{N},\, M,\, R\hspace{-0.5mm}\rightarrow\hspace{-0.5mm}\infty$
that includes $\bar{N}/R\hspace{-0.5mm}=\hspace{-0.5mm}\text{const}$
and $\sqrt{M}/\bar{N}\hspace{-0.5mm}=\hspace{-0.5mm}\text{const}$.
For a fluctuating condensate in the C(I) regime, the time dependent
intensity correlations $g^{(2)}(\tau)\hspace{-0.7mm}=\hspace{-0.7mm}\left\langle n_{0}(t)\,(n_{0}(t\hspace{-0.7mm}+\hspace{-0.7mm}\tau)\hspace{-0.7mm}-\hspace{-0.7mm}1)\right\rangle /\bar{n}_{0}^{2}$
decay like $g^{(2)}(\tau)\hspace{-0.7mm}=\hspace{-0.7mm}1\hspace{-0.7mm}+\hspace{-0.7mm}\exp(-\tau/\tau_{\text{c}}^{(2)})$,
with $\tau_{\text{c}}^{(2)}\hspace{-0.7mm}=\hspace{-0.7mm}\bar{n}_{0}/\hat{B}_{21}X$
as the second order correlation time \cite{Supplemental}, similar
to the photon bunching of thermal emitters.

To conclude we have studied the thermalization and condensation of
photons in equilibrium with a dye microcavity. Our analysis predicts
a regime with unusually large condensate fluctuations, not observed
in present atomic BEC experiments. Moreover, it adds to a classification
of photonic BEC in relation to other light sources. In general, Bose-Einstein
condensation applies to a system in thermodynamic equilibrium; in
contrast to lasing which normally occurs under non-equilibrium conditions.
The latter is brought about by a violation of chemical equilibrium
which occurs if photons are lost (e.g. transmitted) instead of being
reabsorbed by the medium. Additionally, BEC shows a pronounced dependence
on the density of states of a system intrinsically requiring a statistical
multi-mode treatment, whereas single-mode treatments are fully sufficient
to describe the (single-mode) behavior of a wide class of lasers.
If one considers a zero-delay autocorrelation function of $g^{(2)}(0)=1$
to be an essential ingredient of lasing, as it is common in laser
literature \cite{Siegman}, the intensity fluctuations (photon statistics)
can give a further distinguishing feature between lasing and photonic
BEC, as the latter is not necessarily accompanied by a damping of
fluctuations. We note that for other lasing threshold definitions
based on the internal operation of the device, also laser light with
non-poissonian photon statistics can occur \cite{Hofmann,Yien}. This
ambiguity does not occur in the case of photonic BEC, where the phase
transition is uniquely defined.

For the future, it will be important to study the statistics of an
interacting photon gas. In general, interactions can suppress above-poissonian
intensity fluctuations \cite{Kocharovsky}. One here has to distinguish
between thermo-optically induced interactions \cite{Klaers}, which
are comparatively slow, and ultrafast Kerr-lensing. Another line of
research could also be dedicated to the (interaction induced) superfluidity
of the photon condensate. 

We acknowledge financial support of the DFG under contract We1748/17.

\end{document}